# Attention on the Wires (AttWire): A Foundation Model for Detecting Devices and Catheters in X-ray Fluoroscopic Images

YingLiang Ma, Sandra Howell, Aldo Rinaldi, Tarv Dhanjal and Kawal S. Rhode

*Abstract*— *Objective*: Interventional devices, catheters and insertable imaging devices such as transesophageal echo (TOE) probes are routinely used in minimally invasive cardiovascular procedures. Detecting their positions and orientations in X-ray fluoroscopic images is important for many clinical applications. *Method*: In this paper, a novel attention mechanism was designed to guide a convolution neural network (CNN) model to the areas of wires in X-ray images, as nearly all interventional devices and catheters used in cardiovascular procedures contain wires. The attention mechanism includes multi-scale Gaussian derivative filters and a dot-product-based attention layer. By utilizing the proposed attention mechanism, a lightweight foundation model can be created to detect multiple objects simultaneously with higher precision and real-time speed. *Results*: The proposed model was trained and tested on a total of 12,438 X-ray images. An accuracy of 0.88 ± 0.04 was achieved for detecting an echo probe and 0.87 ± 0.11 for detecting an artificial valve at 58 FPS. The accuracy was measured by intersection-over-union (IoU). We also achieved a 99.8% success rate in detecting a 10-electrode catheter and a 97.8% success rate in detecting an ablation catheter. *Conclusion*: Our detection foundation model can simultaneously detect and identify both interventional devices and flexible catheters in real-time X-ray fluoroscopic images. *Significance*: The proposed model employs a novel attention mechanism to achieve high-performance object detection, making it suitable for various clinical applications and robotic-assisted surgeries. Codes are available at https://github.com/YingLiangMa/AttWire.

## I. INTRODUCTION

Minimally invasive cardiovascular procedures are routinely carried out to treat diseases such as coronary heart diseases, valvular heart disease, congenital heart diseases and more. The procedure is usually guided using X-ray fluoroscopy and interventional devices, guidewires, catheters and insertable imaging devices are routinely used during the procedure. Real-time object detection is one of the most important tasks in hybrid guidance systems as well as robotic procedure systems. In hybrid guidance systems for minimally invasive cardiovascular procedures, information is fused from MRI images, CT images or real-time 3D transesophageal echo (TOE) with X-ray fluoroscopy [1][2][3]. Detecting interventional devices or catheters can facilitate motion compensation and automatic registration in MRI or CT based hybrid guidance systems [4][5]. Detecting the location and pose of the TOE probe in live X-ray images can facilitate hybrid guidance system using both 3D echo volumes and X-ray fluoroscopic images [6][7]. Furthermore, knowing the locations of devices may allow surgeries with complete or shared autonomy with robots in the near future.

For catheter detection, early work is focused on detection using pre-defined models. Methods based on active contours [8], shape models, and the Kalman filter [9] were developed in [10], but these approaches are prone to errors caused by image artifacts and the presence of other wire-like objects. Ma et al. [11][12] have developed a detection method based on blob detection. However, these methods only work on catheters with electrodes and cannot detect interventional devices. Recently, learning-based approaches have been developed. Wu et al. [13] utilized a learning-based approach to temporally detect catheter electrodes in fluoroscopy sequences, however it requires manual initialization. Yi et al. [14] developed a catheter and tube detection method based on a scale-recurrent network. However, this method only detects the body of catheters and cannot identify the type of catheters.

The detection of the TOE probe and interventional devices in X-ray images has been previously studied. Existing methods can be divided into two categories: traditional computer vision techniques [7][8] and learning-based methods [14][15][16]. Methods based on traditional computer vision techniques are prone to errors due to image artifacts and the presence of other similar objects. Although learning-based methods have demonstrated a great potential to detect devices robustly, they replied on manual feature selection. Therefore, these methods might not be easily transferable to other target devices and are not capable of detecting multiple devices at the same time.

In recent years, state-of-the-art multiple object detection

Manuscript received June xx, 2024. This work was supported by the Engineering and Physical Sciences Research Council (grant number: EP/X023826/1).

Dr. YingLiang Ma is with School of Computing Sciences, University of East Anglia, United Kingdom (corresponding author: e-mail: yingliang.ma@uea.ac.uk).

Prof. Tarv Dhanjal are with Warwick Medical School, The University of Warwick, Coventry, CV4 7AL, United Kingdom.

Sandra Howell, Prof. Aldo Rinaldi and Prof. Kawal Rhode are with School of Biomedical Engineering and Imaging Sciences, King's College London, St. Thomas' Hospital, London, SE1 7EH, United Kingdom.

methods have been developed to detect and identify common objects such as vehicles, people, and animals [17]. Single-stage detectors such as the YOLO series [18], SSD [19], CenterNet [20], and CornerNet [21] have demonstrated real-time detection speed and the ability to detect multiple objects simultaneously. A comprehensive survey of object detection algorithms in medical images can be found in [22]. Most of these methods use axis-aligned bounding boxes to locate the target objects. A few deep learning-based object detection methods using rotated bounding boxes have been developed, primarily in the domain of satellite image analysis [23].

However, none of the existing deep learning models is capable to detect both interventional devices and flexible catheters. Additionally, these models do not have sufficient accuracy and robustness to be used in our clinical applications. Therefore, in this paper, we designed a foundation model from scratch to take advantage of additional information available in X-ray images. As all interventional devices or catheters are located on a wire and inserted into a major blood vessel and insertable imaging devices are tube-like structures, a novel attention mechanism using trainable pre-defined filters and an attention layer was designed to guide our convolution neural network (CNN) model to the areas of wires and tube-like structures where devices and catheters could be deployed or used. Furthermore, our proposed model uses rotated bounding boxes to detect not only the position of the target object but also its orientation. Since medical devices in X-ray images often have arbitrary orientations, rotated bounding boxes provide a more accurate determination of their locations. As our model uses Gaussian-based center-point heatmap [20] to locate the center of the target object, it enables multiple object detection and localizes the centers of small objects (e.g. electrodes on the catheter). Based on the detected electrode pattern, catheter can be identified [11].

The main contribution of this manuscript is the design of a novel attention mechanism integrated within a point-based object detector to achieve high-performance and multiple object detection in real-time. The proposed foundation model is also capable of detecting and identifying electrode catheters using center-point heatmaps. The model was trained and tested using a large clinical data set.

## II. IMAGE ACQUISITION

12,438 X-ray images were acquired in 72 different clinical cases using two mono-plane X-ray systems (both are Philips X-ray systems) at St. Thomas' Hospital London and University Hospitals Coventry & Warwickshire. There were 6,533 images from 9 transcatheter aortic valve replacement (TAVR) procedures, 250 images from one atrial fibrillation (AF) ablation procedure guided by X-ray and transesophageal echo (TOE) images and 5,655 images from 62 standard ablation procedures.

To facilitate image synthesis, the TOE probe was scanned using a research X-ray system at King's College London. The C-arm was rotated from LAO 90 degrees to RAO 90 degrees in 2-degree increments, resulting in the acquisition of 90 X-ray images. These images were later used to create synthetic images. The X-ray system used for acquisition was the Siemens Axiom Artis Cath System.

## III. METHODS

### A. Image synthesis

As 5,655 images out of total 12,438 images do not contain the TOE probe, a method of image synthesis has been developed to automatically insert an image patch of a TOE probe. It is based on Poisson image editing (PIE) [24][25], which blends an image patch into the context of a destination image. The blending was achieved via solving the equation (1).

$$\min_{f_{in}} \iint_\Omega |\nabla f_{in} - v|^2 \text{ with } f_{in}|_{\partial\Omega} = f_{out}|_{\partial\Omega} \qquad (1)$$

where $\nabla$ is the gradient operator. The goal of eq. (1) is to find the intensity values $f_{in}$ within the masked area ($\Omega$) of image patch matching with the surrounding values $f_{out}$ of the destination image. A binary mask will be used to create the masked area ($\Omega$), which is the loose selection of the blending object. $\partial\Omega$ is the border of the masked area and $v$ is the image gradient within the masked area. Figure 1 gives an example of PIE.

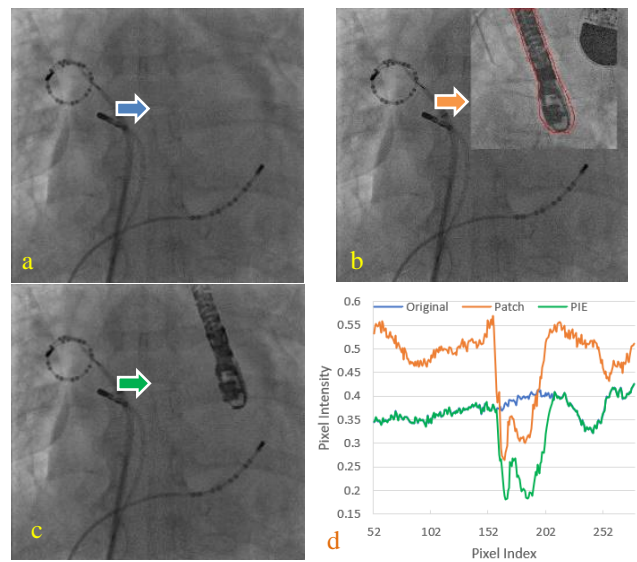

Fig. 1. An example of PIE. (a) The original image. (b) Overlay the image patch with the original image. The red contour is the border of the masked area ($\partial\Omega$). (c) Image after applying PIE. (d) The intensity profiles. Arrows in the images indicate the location of the intensity profiles.

90 image patches were extracted from 90 X-ray images which contain the TOE probe and were acquired in the research X-ray system. Data augmentation techniques were used to increase the variation of the pose of the TOE probe in X-ray images. Random rotations and translations were applied to the extracted image patches. There are restrictions applied to random rotations and translations to ensure the generated image is anatomically correct.

As the image patch is rotated by a random angle, it is essential to calculate the largest rectangle in a rotated rectangle to crop the rotated image. As shown in figure 2b, to simplify the problem, the largest rectangle (yellow) with the same aspect ratio as the minimum bounding rectangle (black) is chosen as our cropping rectangle and it is located on diagonal lines. To compute the size, the width of rotated image can first be computed as $w_{rot} = w\cos\alpha + h\sin\alpha$ (see figure 2c). Similarly, the height can be calculated as $h_{rot} = w\sin\alpha + h\cos\alpha$. As shown in figure 2d, $\beta = tan^{-1}(w_{rot}/h_{rot})$ and the length $\|BC\|=h\cdot\cos\alpha\cdot\sin\alpha$. The length BC can be also computed as $\|BC\| = \|BD\|\sin\gamma$, where $\gamma = \alpha + \beta$. Therefore,

$$\|BD\| = h \cdot \cos\alpha \cdot \sin\alpha / \sin(\alpha + \beta) \quad (2)$$

Finally, the offset $dx = \|BD\|\sin\beta$ and offset $dy = \|BD\|\cos\beta$. Applying data augmentation techniques for the image patches not only provides more training data, but also enables automatic data labelling as the same affine transformation matrices can be applied to the original rotated bounding box to generate the remaining bounding boxes.

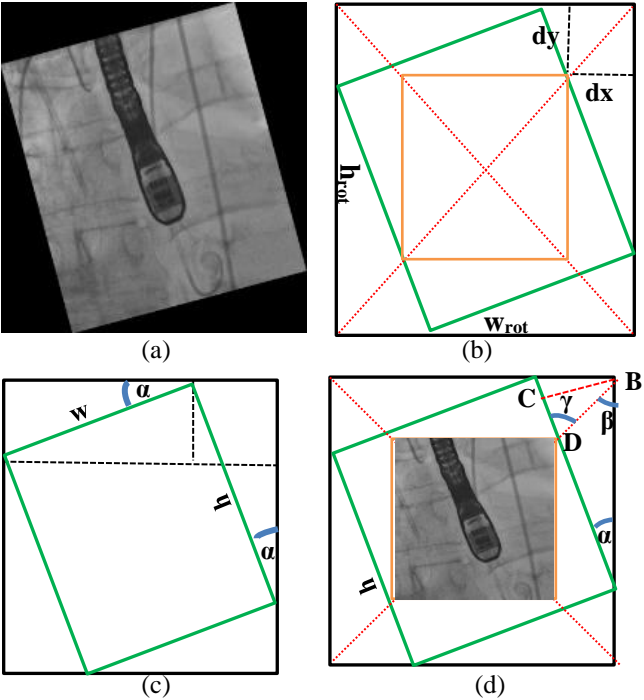

Fig. 2. The largest rectangle in a rotated rectangle. (a) The rotated patch. (b)(c)(d) Illustrations of computational steps.

### B. Attention on the wires (AttWire): the attention backbone.

Our clinical applications require real-time object detection while maintaining high accuracy and robustness. To achieve this goal, attention mechanisms were designed to take advantage of additional information about the location and structure of medical devices. Many devices contain wire and wire mesh or tube-like structures. To guide the attention of our CNN models, the multi-scale Gaussian derivative filters were used in the first convolution layer to enhance the visibility of wire-like or tube-like objects [12]. This process involves the calculation of a 2x2 Hessian matrix, and it is computed at every image pixel [12]. The Hessian matrix H consists of second order derivatives that contain information about the local curvature. H is defined such as:

$$H = \begin{bmatrix} L_{xx}(x,y;s) & L_{xy}(x,y;s) \\ L_{yx}(x,y;s) & L_{yy}(x,y;s) \end{bmatrix} \quad (3)$$

Where $G_{xx}(x,y;s)$, $G_{yy}(x,y;s)$ and $G_{xy}(x,y;s)$ are Gaussian derivatives (DoG). In practice, we just pre-compute the masks of these Gaussian derivatives, convolve with the input image. By combining these multi-scale Gaussian derivative filters together, they can provide a global attention on the wire-like or tube-like structures.

The architecture of the attention backbone is illustrated in figure 3 and 15 DoG filters are used in the first convolution layer to provide the first attention component. Among 15 filters, there are five groups, and each group contains three DoG filters with the same scale factor s, which are defined as $G_{xx}(x,y;s)$, $G_{yy}(x,y;s)$ or $G_{xy}(x,y;s)$. To accommodate different sizes of objects on the wires, five different scale factors were used in five groups of LoG filters. To calculate the scale factor $s_0$ for object size $r_0$, we use $s_0 = ((r_0 - 1)/3)^2$. This equation is motivated by the "$3\sigma$" ($s_0 = \sigma^2$) rule that 99% of energy of the Gaussian is within three standard deviations. The final multiscale $s_0$ is in the range of $0.11 \leq s_0 \leq 9$ and it is based object size from 2 to 10 (Unit is in image pixels) in an image with a 200x200 resolution. Although these filters have been pre-defined and have preset parameters, they are trainable during the learning process. Therefore, the attention mechanism can be adapted to learn to detect different objects with different sizes. The second attention component is achieved by a dot-product based attention layer [26], which calculates the similarity between the random filter output and the output from LoG filters (figure 3). The attention layer further enhances the attention on the wire-like or tube-like structures.

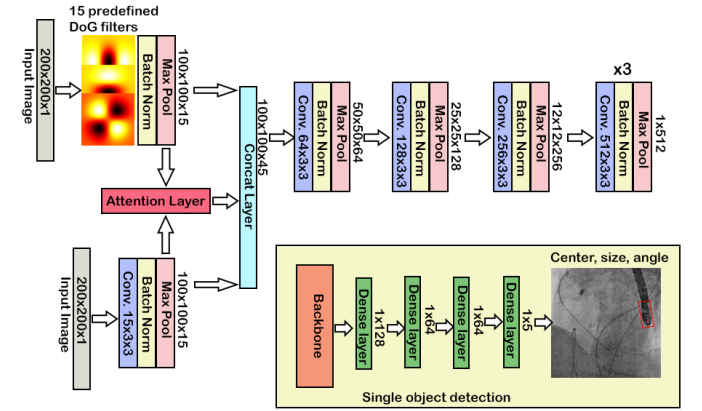

Fig. 3. Attention backbone and an example of its usage in single object detection.

### C. Single object detection.

We first designed a customized CNN with the proposed attention backbone for single object detection to test its performance. As shown in figure 3, the localization of a rotated bounding box is achieved by the output of the final dense layer, which provides five parameters: center *(x, y)*, size *(w, h)* and angle (δ).

Two object detectors were trained, one for the TOE probe and the other one for the transcatheter aortic valve (before deployment). A modulated rotation loss function was designed, and it minimizes the difference between the predicted values and ground truth values. All five parameters which define the rotated bonding box were normalized between 0 and 1 to avoid errors caused by objects on different scales. The loss function $l_{mr}$ is defined as:

$$l_{cp} = |x_g - x_p|/W_{img} + |y_g - y_p|/H_{img} \qquad (5)$$

$$\begin{cases} l_1 = l_{cp} + |w_g - w_p|/W_{img} + |h_g - h_p|/H_{img} \\ l_2 = l_{cp} + |w_g - h_p|/W_{img} + |h_g - w_p|/H_{img} \end{cases} \qquad (6)$$

$$l_{mr} = min \begin{cases} l_1 + |\delta_g - \delta_p|/90° \\ l_2 + |90° - |\delta_g - \delta_p||/90° \end{cases} \qquad (7)$$

where $l_{cp}$ is the central point loss, $(x_p, y_p)$ is the predicted center point and $(x_g, y_g)$ is the ground-truth center point. Eq. (6) and (7) are for the exchangeability of height and width.

As shown in figure 4, the activation maps of selected layers were visualized to illustrate the model attentions in the aortic valve detector. The model global attention is clearly on the wire-like structures in the first layer and then enhanced by the attention layer. Finally, the model shifts the attention to the local areas of the target object (the aortic valve) in the final convolution layer.

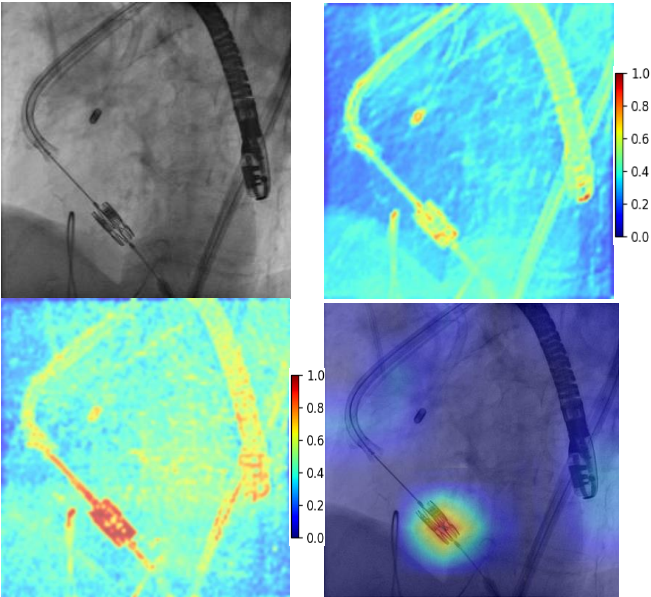

Fig. 4. (a) The original image. (b) The activation map from the convolution layer with 15 DoG filters. (c) The activation map from the attention layer. (d) The activation map from the last convolution layer in the attention backbone.

### D. Multiple object detection

Inspired by the CenterNet [20], a one-stage multiple-object detector was designed by using a similar attention backbone proposed in this paper. The one-stage detector can achieve a higher inference speed and it is suitable for real-time applications. The proposed detector is a light-weight CNN model and only contains 3.7M trainable parameters for three-class object detector. As shown in figure 5, the proposed CNN model consists of down-sampling layers and up-sampling layers. The model has four outputs. The first one is the center-point heatmap and it is used to localize the center point *(x, y)* of the rotated bounding box. The second output is used to determine the object size *(w, h)*. The third output is the rotate angle (δ) of the bounding box. The fourth output is the offset output, and it is used to recover from the discretization error caused due to the downsampling of the input. For example, in our model, the input image resolution is *W*x*H* and the image resolution after the last up-sampling layer is $W/2 \times H/2$. If the ground truth of center point is $(x_g, y_g)$ in center heatmap output, the corresponding ground truth of center point in the input image is $(2x_g + \varepsilon_x, 2y_g + \varepsilon_y)$. Both $\varepsilon_x$ and $\varepsilon_y$ are discretization errors and they are either 0 or 1 in our model.

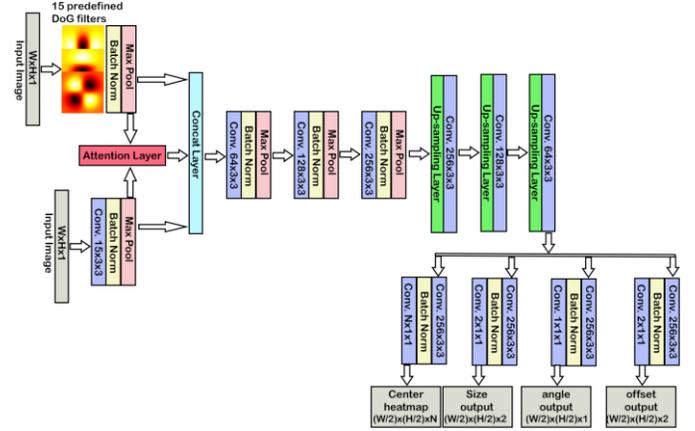

Fig. 5. The CNN architecture for multiple object detection. N is the number of classes. The resolution of input image is *W*x*H* and the resolution of output images is (*W/2*)x(*H/2*).

The proposed CNN model not only outputs a rotated bounding box for each object but also outputs a confidence value *CL*. *CL* is defined as the local maximum value in each center-point heatmap. When *CL>0.5*, a class of objects is detected. The threshold of 0.5 was chosen because it is the midpoint between 0 and 1. A value of 0 indicates no object is present, while a value of 1 indicates the maximum probability of an existing target object.

Therefore, the model can predict whether the target object exists in the image or not. The model also can achieve multiple object detection as it has multiple channels of center heatmaps and each channel can localize the center points of one class of objects. The ground-truth heatmap for center points is not defined as either 0 or 1 because locations near the target point should get less penalization than locations far away. Therefore, Gaussian heatmap $e^{\frac{\|P-P_g\|^2}{2\sigma^2}}$ was used and *P* is the predicted center point and $P_g$ is the ground truth. $\sigma$ is set to 1/3 of the radius, which is determined by the size of objects. Focal loss [19] is used in the output for center-point heatmap and it is mainly to solve the problem of imbalanced classification in target detection. The loss functions for the remaining outputs are L1 loss function. Figure 6 presents some results of center-point heatmaps and detection results.

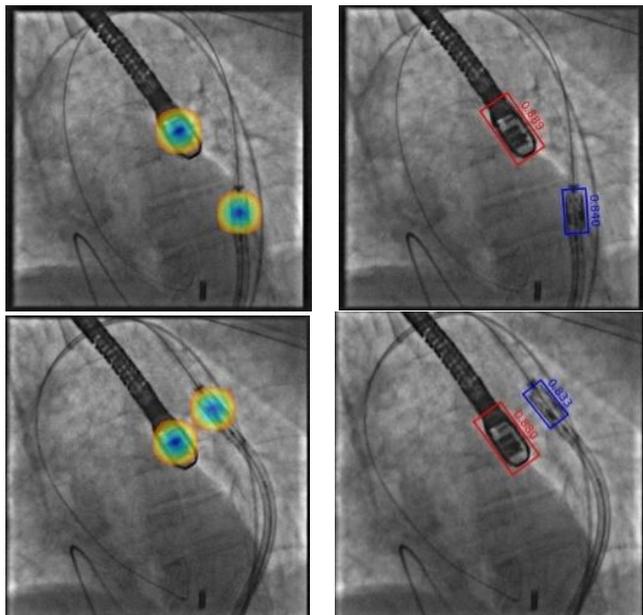

Fig. 6. Center point heatmaps and object detection with confidence values. The red box indicates the location of the TOE probe and the blue box indicates the location of transcatheter aortic valves.

### E. Catheter detection

To detect and identify electrode catheters, we first need to locate the positions of the electrodes on the catheters. We achieve this by using the third channel of the center heatmap output for electrode detection. In our current model, the first two channels are allocated to detecting the centers of interventional devices, such as the TOE probe head, transcatheter aortic valves, transeptal needle, pacing wires and stents. The number of channels can be adjusted based on the specific clinical application, increasing or decreasing as needed. We also use the one of size outputs to determine the size of electrode which is essential for detecting ablation catheters. Figure 7 illustrates how the output channels connect to the detected object centers, object size, and angle information.

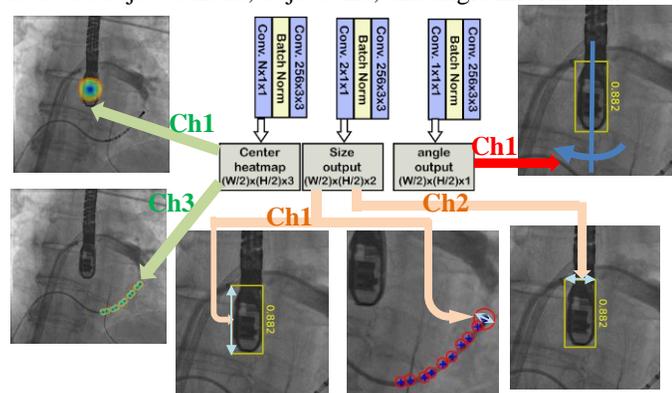

Fig. 7. Output channels from a multiple object detector. There are three channels (green) from the output of center heatmap. Channel 2 has outputted a confidence value of less than 0.5 so no valve is detected. There two channels from the size output. One is for the height of the TOE probe and the size of electrodes. The other one is for the width of the TOE probe. Finally, there is only one channel from the angle output.

To identify an electrode catheter, the pattern of electrodes is analyzed, including the number of electrodes, the distances between adjacent electrodes, and the inner angles formed by neighboring electrodes. The details of catheter identification algorithms can be found in [11] and [12]. Figure 8 presents the results of both the detection of TOE probe and catheter identification. The proposed detector can also be used exclusively to detect various types of electrode catheters (e.g., ablation catheters, ten-electrode decapolar catheters, and defibrillation catheters). In this clinical application, the first two channels of the center heatmap output in our model are not utilized, as there is no TOE probe, valve or other devices. Figure 9 presents some examples.

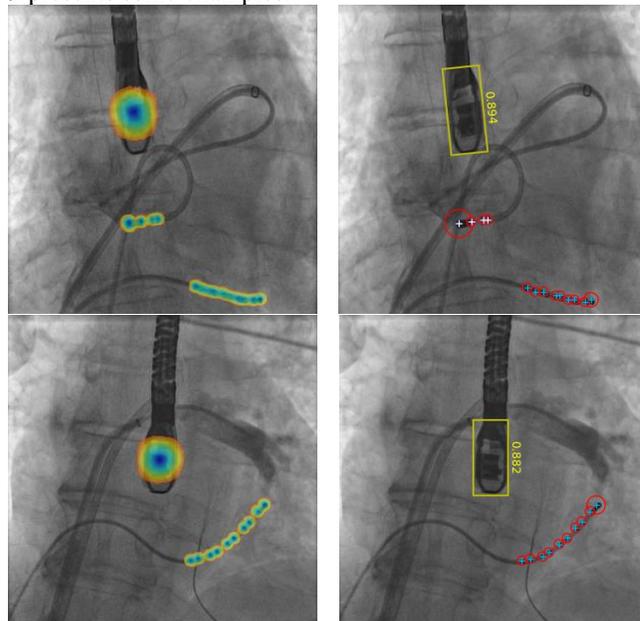

Fig. 8. Center heatmaps (left) and the detection of TOE probe and catheters (right).

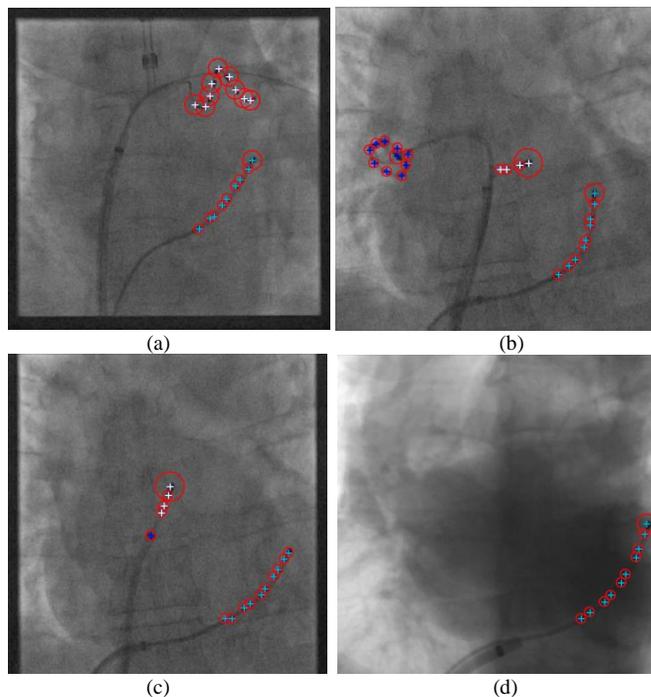

Fig. 9. Examples of detecting catheters only. (a) White crosses are detected defibrillation catheters and light blue crosses are detected 10- electrode catheter. (b)(c)(d) White crosses are detected ablation catheters and light blue crosses are detected 10-electrode catheter. Blue crosses are detected other objects.

## IV. RESULTS

A total of 12,438 X-ray images were used to train and test all object detectors, including 5,655 synthetic images. The datasets were split with a ratio of 80:10:10 for training, validation, and testing. Both validation and testing images are real images, and 5,655 synthetic images were only used in the training dataset. Ninety X-ray images from the rotational scan of the TOE probe head were used exclusively for image synthesis and were not included in the training, validation, or testing datasets. All our deep learning models for object detection using different backbones were implemented in Keras with a Tensorflow (version 2.10) backend and were trained on a GPU farm (NVidia RTX 6000 Ada with 24G memory). The trained models were evaluated on an Intel i7 1.8GHz laptop with a NVidia T550 graphics card to test the inference speed.

### A. Evaluation of single object detection

We implemented our CNN model for single-object detection using VGG16 and ResNet-50 backbones, along with the proposed attention backbone (AttWire), to evaluate their performance. The network architecture is shown in Figure 3. We used the Keras implementations of the VGG16 and ResNet-50 backbones with pre-trained ImageNet weights. All models utilized the Adam optimizer, with the learning rate set to 0.001. The activation function is Rectified Linear Unit (ReLU). The loss function is defined in Eq(7). All models were trained for 100 epochs and use the batch size of 8.

Table 1 shows the comparison results of our approach (AttWire) with VGG16 and ResNet-50 backbones in single object detection. $AP_{50}$ and $AP_{75}$ are the average precisions, which are evaluated at $T(IoU)=0.5$ and $T(IoU)=0.75$. mAP is the mean value across different IoU thresholds (IoU thresholds from 0.5 to 0.95 with a step size of 0.05).

Our single object detection model for TOE probe head has achieved 1.0 in $AP_{50}$ and 0.93 in $AP_{75}$ and operates at a speed of over 50 FPS. Therefore, it can facilitate real-time fusion between X-ray fluoroscopy and 3D echo images by providing the in-plane translation and rotation parameters of the TOE probe head within the X-ray fluoroscopic images.

### B. Evaluation of multiple object detection

Our model for multiple object detection is designed not only to detect devices but also to identify the centroids of small objects such as catheter electrodes. It belongs to the family of point-based single-stage object detectors. We believed that other popular single-stage object detectors such as YOLO series and SSD cannot accurately locate the centroids of objects. Therefore, we mainly focus on comparing CenterNet-like object detectors with different backbones and we only compare the performance of our models with state-of-the-art YOLO v8.

Table 2 shows the comparison results of our CNN model for multiple object detection using the proposed attention backbone (AttWire) against CNN models with MobileNet, DenseNet, and ResNet-50 backbones. The architecture of the CNN models with pre-trained backbone was illustrated in figure 10. All models utilized the Adam optimizer, with the learning rate set to 0.001. The activation function is Rectified Linear Unit (ReLU). Focal loss is used in the output for center-point heatmap and the loss functions for the remaining outputs are L1 loss function. All models were trained for 50 epochs and use the batch size of 8. The resolution of the input image is 300x300. The error in electrode detection is defined as the 2D distance between the manually defined positions and the positions detected by our method. We use the YOLOv8 model from [27] for our performance testing, and the model training uses the same hyperparameters as our CNN models.

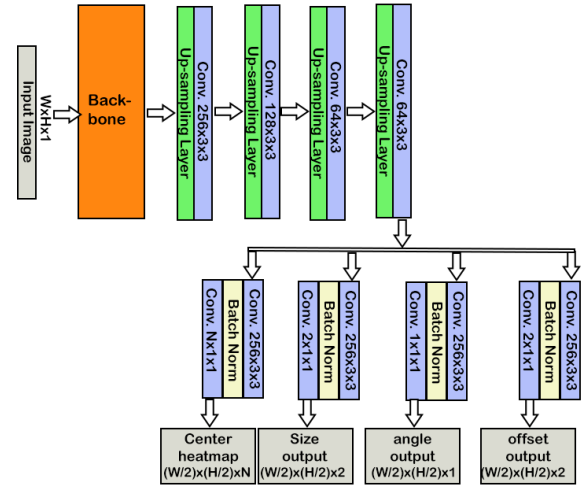

Fig. 10. The architecture of the CNN models with a pre-trained backbone

$AP_{50}$ and $AP_{75}$ in multiple object detection are the mean values of all objects (both TOE probe head and aortic valve). mAP is the mean value across different IoU thresholds (IoU thresholds from 0.5 to 0.95 with a step size of 0.05).

| Target Object | Backbone | Parameters | Speed | IoU | mAP | $AP_{50}$ | $AP_{75}$ |
|---|---|---|---|---|---|---|---|
| TOE probe head | VGG16 | 17.1M | 43 FPS | 0.77±0.21 | 0.656 | 0.9 | 0.812 |
| | ResNet-50 | 36.4M | 31 FPS | 0.84±0.14 | 0.733 | 0.977 | 0.874 |
| | **AttWire** | **6.8M** | **55 FPS** | **0.89±0.05** | **0.815** | **1.0** | **0.928** |
| Aortic valve | VGG16 | 17.1M | 52 FPS | 0.81±0.11 | 0.725 | 0.972 | 0.743 |
| | ResNet-50 | 36.4M | 37 FPS | 0.85±0.08 | 0.808 | 0.981 | 0.896 |
| | **AttWire** | **6.8M** | **59 FPS** | **0.88±0.07** | **0.832** | **1.0** | **0.943** |

Table 1. Results for single object detection.

| Backbone/Detector | Parameters | Speed | IoU (TOE) | IoU (valve) | Electrodes (pixels) | Electrodes (mm) | mAP | AP50 | AP75 |
|---|---|---|---|---|---|---|---|---|---|
| MobileNet | 7.3M | 53 FPS | 0.79±0.09 | 0.77±0.17 | 2.97±1.84 | 1.0±0.63 | 0.546 | 0.999 | 0.603 |
| DenseNet121 | 11.1M | 34 FPS | 0.81±0.07 | 0.79±0.16 | 2.15±1.27 | 0.73±0.43 | 0.584 | 0.997 | 0.644 |
| ResNet-50 | 28.7M | 41 FPS | 0.81±0.07 | 0.80±0.15 | 1.64±1.1 | 0.56±0.37 | 0.728 | 0.998 | 0.759 |
| YOLOv8s (Detector) | 11.4 M | **81 FPS** | 0.84±0.06 | 0.80±0.13 | N/A | N/A | 0.753 | 1.0 | 0.814 |
| **AttWire** | **3.7M** | 58 FPS | **0.88±0.04** | **0.87±0.11** | **1.55±0.8** | **0.53±0.27** | **0.789** | **1.0** | **0.922** |

Table 2. Results for multiple object detection.

## C. Catheter detection results

In this experiment, detection methods for both the ablation catheter and the 10-electrode catheter were evaluated. The error in catheter detection is defined as the average detection error across all electrodes on the catheter. The success rate is defined as: *success rate = 1.0 – failure rate*. For the technique to be acceptable in clinical practice, a failed catheter detection was defined as any instance where any of the electrodes on the catheter had an error exceeding a preset threshold. For example, a threshold of 2.5 mm (approximately 6 image pixels) [11] was used in this evaluation, corresponding to the size of the smallest target structures in cardiac catheterization procedures. Table 3 show the comparison results of our CNN model-based approach and blob detection-based method [10].

| Methods | Ablation catheter (mm) | success rate | 10-electrode catheter (mm) | success rate |
|---|---|---|---|---|
| Blob detect method [10] | 0.98 ± 0.73 | 95.7% | 0.65 ± 0.46 | 97.4% |
| **AttWire** | **0.91 ± 0.68** | **97.8%** | **0.55 ± 0.34** | **99.8%** |

Table 3. Catheter detection errors for both ablation catheter and 10-electrode catheter.

## D. Ablation studies

The ablation studies on different modification of our CNN model for multiple object detection using AttWire is necessary to investigate the features responsible for the detection performance improvement. Table 4 and 5 clearly demonstrate that without the attention components, the model's performance is significantly worse than that of the complete model. Furthermore, removing the attention components has less impact on the multiple object detection model than on the single object detection model, as the CenterNet-like CNN model has superior performance in localizing the center of the target object.

| Target | Model components | IoU | AP$_{50}$ | AP$_{75}$ |
|---|---|---|---|---|
| TOE probe head | Remove LoG filter and attention layers | 0.78±0.16 | 0.947 | 0.824 |
| | Remove attention layer | 0.83±0.09 | 0.997 | 0.858 |
| | Completed AttWire Model | 0.89±0.05 | 1.0 | 0.928 |
| Aortic valve | Remove LoG filter and attention layers | 0.80±0.13 | 0.968 | 0.761 |
| | Remove attention layer | 0.82±0.09 | 0.992 | 0.823 |
| | Completed AttWire Model | 0.88±0.07 | 1.0 | 0.943 |

Table 4. Results for single object detection.

| Model components | IoU (TOE) | IoU (valve) | AP$_{50}$ | AP$_{75}$ |
|---|---|---|---|---|
| Remove LoG filter and attention layers | 0.83±0.07 | 0.80±0.17 | 0.996 | 0.79 |
| Remove attention layer | 0.85±0.06 | 0.81±0.15 | 1.0 | 0.82 |
| Completed AttWire Model | 0.88±0.04 | 0.87±0.11 | 1.0 | 0.92 |

Table 5. Results for multiple object detection.

## V. DISCUSSION AND CONCLUSION

Clinical applications for minimally invasive heart procedures require highly robust and accurate algorithms for detecting interventional and imaging devices in real-time X-ray fluoroscopic images. In this paper, novel attention mechanisms were designed to guide the CNN model to the areas of wires in X-ray images. The attention-based backbones were implemented in our CNN models for both single and multiple object detection, outperforming existing state-of-the-art and lightweight backbones across every metric. Additionally, our model for multiple object detection outperforms the state-of-the-art YOLO v8 model in most metrics, except for detection speed. Therefore, it could facilitate real-time fusion between X-ray fluoroscopy and 3D echo images by providing the in-plane translation and rotation parameters for the TOE probe head in real-time. By detecting additional objects (e.g., transcatheter aortic valve) simultaneously, our foundation model could provide the completed guidance solution for transcatheter aortic valve (TAVI) implantation procedures. It will combine the advantages of both image modalities as X-ray images provide the accurate location of artificial aortic valve and ultrasound images provide the implantation location.

Another advantage of the proposed model is its ability to detect multiple catheters by localizing the electrode positions. By analyzing the pattern of catheter electrode, each catheter could be identified. Therefore, our foundation model is capable of detecting both catheters and other interventional devices simultaneously. Furthermore, our attention backbone (AttWire) could be integrated into a standard U-Net, enabling high-performance segmentation of guidewires and the full body of electrode catheters. We have conducted some preliminary studies using a standard U-Net for segmenting pacing leads and wires in [28].

Our foundation model for multiple object detection is highly adaptable, capable of detecting landmarks on guidewires or pacing leads. This flexibility paves the way for numerous clinical applications in guiding cardiac interventional procedures. For instance, detecting the catheter inside the coronary sinus could facilitate real-time motion compensation for overlaying 3D anatomical model with X-ray fluoroscopic images. With the rise of robot-assisted interventions, automatic tracking of catheters and interventional devices has become a crucial component of robotic systems. This capability will enable procedures to be performed with full or partial autonomy by robots in the near future. A potential improvement is to incorporate a temporal attention component into our model for tracking objects in video sequences, enhancing detection consistency between adjacent frames.

## VI. ACKNOWLEDGEMENT

This work was supported by the Engineering and Physical Sciences Research Council (grant number: EP/X023826/1); and the National Institute for Health Research (NIHR) Biomedical Research Centre at Guy's and St Thomas' NHS Foundation Trust and King's College London. The views expressed are those of the author(s) and not necessarily those of the NHS, the NIHR or the Department of Health.